\documentclass[epj]{svjour}
\usepackage{graphics}
\begin{document}
\title{Adaptive social recommendation in a multiple category landscape}
\author{Duanbing Chen \inst{1} \and An Zeng\inst{2}\thanks{\email{an.zeng@unifr.ch}} \and Giulio Cimini\inst{2} \and Yi-Cheng Zhang\inst{2}}
\institute{\inst{1} Web Sciences Center, University of Electronic Science and Technology of China, Chengdu 611731, People's Republic of China\\
  \inst{2} Physics Department, University of Fribourg, Chemin du Mus\'{e}e 3, CH-1700 Fribourg, Switzerland\\}
\date{Received: date / Revised version: date}
%
\abstract{
People in the Internet era have to cope with the information overload, striving to find what they are interested in,
and usually face this situation by following a limited number of sources or friends that best match their interests.
A recent line of research, namely adaptive social recommendation, has therefore emerged to optimize the information propagation in social networks and provide users with personalized recommendations. Validation of these methods by agent-based simulations often assumes that the tastes of users and can be represented by binary vectors, with entries denoting users' preferences. In this work we introduce a more realistic assumption that users' tastes are modeled by multiple vectors.
We show that within this framework the social recommendation process has a poor outcome. Accordingly, we design novel measures of users' taste similarity
that can substantially improve the precision of the recommender system. Finally, we discuss the issue of enhancing the recommendations' diversity while preserving their accuracy.
} 
\maketitle
\section{Introduction}

We live in the information and communications technology (ICT) based society where information is overabundant, and where recommender systems are widely used to filter out irrelevant information~\cite{PR5191}.
Common techniques to obtain recommendations include collaborative filtering~\cite{ACMTIS225,IEEEIC776}, Bayesian clustering~\cite{PCUAI1998}, probabilistic latent semantic analysis~\cite{ACMTIS2289},
matrix decomposition~\cite{PRL87248701}, mass diffusion~\cite{PRE76046115} and heat conduction~\cite{PRL99154301}. Many issues related to recommender systems have also been considered,
such as the diversity of the recommendations~\cite{PNAS1074511}, the influence of the network topology~\cite{PhysicaA3911822} and the feedback effect of iterated recommendations~\cite{EPL9718005}.

Recently, the advent of information-sharing websites like Twitter, Facebook and Digg, where users select others as information sources or friends and import stories or posts from them,
has shifted the paradigm of recommender systems to the social ground. Specifically, an approach named social recommendation has emerged to make direct use of the connections
between the members of a society~\cite{Golbeck-Science-2008}. The outcome of such recommendation process thus depends on the structure of the network of connections,
with higher success rate if linked users share similar interests.

A newly proposed adaptive method for social recommendation~\cite{Medo-2009} is based on the process of information diffusion in a social system
where connections evolve (adapt) in order to link users with similar interests. In other words, the system analyses the information consumption patterns of users
and assigns to each of them suitable information sources. Then users obtain recommendations as a natural result of the information spreading process.
The model has been extensively tested by agent-based simulations~\cite{Medo-2009}, and additional aspects like users' reputation~\cite{Giulio-2011}, implicit ratings~\cite{Wei-2011}, local topology optimization~\cite{Chen-2011},
leadership structure~\cite{Plosone620648} and link reciprocity~\cite{Chen-2012} were subsequently investigated in detail.

The agent-based framework used in these works assumes that users' interests are modeled by binary taste vectors, with entries indicating whether a user has preference for some category
(which can be music, movies, science, politics, to name a few). This assumption may appear to be too simplistic to model real users, as each category can have sub-categories:
there are different music genders, kind of movies, scientific disciplines, and so on, and an user who likes classical music may be not interested in or even totally dislike rap.
Additionally, real users are heterogeneous in the number of categories and topics they like, for instance user $i$ can be interested in science, politics and sport, whereas user $j$ may like music and movies.

In this work we build on a more realistic assumption: users' interests are represented by multiple vectors, meaning that each category is represented by a binary vector
with entries giving the preference for the relative sub-categories. We test the robustness of the adaptive recommendation method within this assumption,
and identify the measures of users' taste similarity which are efficient for constructing the social network to obtain accurate recommendations. The properties of the network are also analysed in detail.
Finally we propose a method to considerably enhance the diversity of the recommendation process, while preserving its accuracy.

\section{Model description}

We first briefly summarize the adaptive recommendation method introduced in~\cite{Medo-2009}.
The system consists of $U$ users, each is connected by directed links to $L$ other users, who represents her information sources and to whom we refer as her \emph{leaders}.
The value of $L$ is fixed as users can follow a limited number of sources.
Users receive pieces of information (we will speak about news for brevity) from their leaders, and eventually assess them. In addition, they can introduce new content to the system.
Evaluation of news $\alpha$ by user $i$ ($e_{i\alpha}$) is either $+1$ (liked), $-1$ (disliked) or $0$ (not read yet).
The set of evaluations from any pair of users $i$ and $j$ is the basis to compute their similarity of interests (or reading tastes), which we denote as $s_{ij}$.
The explicit recipes to compute users' similarity are presented in the next section. We remark that, apart from their evaluations, no other information about users is assumed by the model.

\subsection{Propagation of news}
When news $\alpha$ is introduced to the system by user $i$ at time $t_{\alpha}$, it is passed from $i$ to the users $j$ who have selected her as a leader (to whom we refer as her \emph{followers}),
with a \emph{recommendation score} proportional to their similarity $s_{ij}$. If this news is later liked by one of users $j$ who received it, it is similarly passed further to this user's followers $k$,
with recommendation score proportional to $s_{jk}$, and so on. Summarizing, for a generic user $k$ at time $t$, a news $\alpha$ is recommended to her according to its current score:
\begin{equation}\label{eqRecScore}
R_{k\alpha}(t)=\delta_{e_{k\alpha},0}\:\lambda^{t-t_{\alpha}}\,\sum_{l\in L_k}s_{kl}\:\delta_{e_{l\alpha},1}
\end{equation}
where $L_k$ is the set of leaders of user $k$, the term $\delta_{e_{k\alpha},0}$ equals one only when user $k$ has not read news $\alpha$ yet and the term $\delta_{e_{l\alpha},1}$
is one only if user $l$ liked news $\alpha$. To allow fresh news to be accessed fast, recommendation scores are also damped with time ($\lambda\in(0,1]$ is the damping factor).

\subsection{Leader selection}

As the model is adaptive, leader-follower connections are periodically rewired to have the social network in an optimal state where users with high similarity are directly connected.
When rewiring occurs for user $i$, her current leader with the lowest similarity value ($j$) is replaced with a new user ($k$) if $s_{ik}>s_{ij}$.
There are different selection strategies for picking new candidate leaders, which are discussed in detail in~\cite{Giulio-2011,Chen-2011}.
In this work we employ a hybrid strategy for which the user $k$ is picked at random in the network with probability 0.1,
otherwise she is selected among the leaders' leaders and followers of user $i$ to maximize $s_{ik}$.
This mechanism well mimics users establishing mutual friendship relations, searching for friends among friends of friends, and having casual encounters which may lead to long-term relationships.
In addition, it is an excellent compromise between computational cost and system's performance~\cite{Chen-2011}.

\section{Measure of users' similarity}

It is clear from the previous section that users' similarity is a crucial ingredient of the model, as it determines both recommendation scores and the leader selection process.
For the recommender system to work is hence important to have reliable similarity measures, which however can only be estimated by comparing users' past assessments.

The definition of the similarity used in~\cite{Medo-2009} is based on the overall probability of agreement: for a pair of users $i$ and $j$,
\begin{equation}\label{eqSimO}
s_{ij}^{(0)}=\frac{|A_i \bigcap A_j|+|D_i \bigcap D_j|}{|N_{i}\bigcap N_{j}|}\left(1-\frac1{\sqrt{|N_{i}\bigcap N_{j}|}}\right)
\end{equation}
where $A_i$ and $D_i$ ($A_j$ and $D_j$) are respectively the set of news approved and disapproved by user $i$ (by user $j$),
and $N_{i}$ ($N_{j}$) is the set of news read by user $i$ (by user $j$), with $N_{i}=A_{i}\bigcup D_{i}$ ($N_{j}=A_{j}\bigcup D_{j}$).
The term in the parentheses is intended to disfavor user pairs with small overlap, which are more sensitive to statistical fluctuations.

The similarity measure just introduced is symmetric: $s^{(0)}_{ij}\equiv s^{(0)}_{ji}$. However the leader-follower relation is not symmetric,
as news propagate from leader to follower and not viceversa---unless the link is reciprocal. It is in fact often the case that user $j$ can be a good leader for user $i$, whereas the opposite does not hold.
For instance, $j$ may be interested only in a few categories like music and sport, whereas $i$ may have much broader interests (music, sport, politics, economics).
In this case, $j$ is more selective in news' evaluation than $i$, and if she is selected by $i$ as a leader, she will forward only the news belonging to her few favorite categories---which also match $i$'s interests.
If instead $j$ selects $i$ as a leader, she will receive more diverse news, including the ones in which she is not interested.
For the sake of users' satisfaction, $j$ should be the leader and $i$ the follower, meaning that $s_{ij}\gg s_{ji}$. According to these considerations, we modify $s_{ij}^{(0)}$ to build an asymmetric similarity measure as:
\begin{equation}\label{eqS1}
s_{ij}^{(1)}=\frac{|A_i \bigcap A_j|+|D_i \bigcap D_j|}{|N_{j}|}\left(1-\frac1{\sqrt{|N_{j}|}}\right)
\end{equation}
which is the probability of agreement on the set of news assessed by $j$ (the actual or candidate leader).

Two remarks are in order at this point. When the total number of categories is big, users are usually interested in only a limited number of them,
and the fact that any two users disapprove many news in common means that their favorite categories do not overlap, but it does not imply that they are similar.
Hence the term $|D_i \bigcap D_j|$ in $s^{(1)}$ can be misleading. Additionally, when assessing the quality of a leader, it would be more appropriate to refer only to the news liked by the leader,
which are the ones that are actually passed to and eventually evaluated by her followers. Consequently, we further introduced another similarity index:
\begin{equation}\label{eqS2}
s_{ij}^{(2)}=\frac{|A_i \bigcap A_j|}{|A_j|}\left(1-\frac1{\sqrt{|A_j|}}\right)
\end{equation}
which, by not considering dislikes, is basically a Jaccard coefficient representing the probability of $i$ liking a news liked by $j$.

As a final remark, we note that a good leader not only has to forward news that are liked by her followers, but also has to block the news that they might dislike.
We can hence introduce another term in definition (\ref{eqS2}) in order to minimize the probability of $i$ disliking a news liked and forwarded by $j$. We obtain:
\begin{equation}\label{eqSimN}
s_{ij}^{(3)}=\frac{|A_i \bigcap A_j|-|D_i\bigcap A_j|}{|A_j|}\left(1-\frac{1}{\sqrt{|A_j|}}\right)
\end{equation}

Note that in all the definitions above, when the similarity is undefined it is replaced by a small value $s_0$.

In what follows, we will study the behavior of the system under these similarity metrics.
For numerical tests of the model, we use a new agent-based framework.

\section{Agent-based simulations}

To model users' interests and news' attributes we use a multiple vector model.
There are a total of $M$ different categories of news in the system (for instance: music, movies, science, politics, business, technology, sport, gossip, and so on).
A generic user $i$ has preference for $1\le m_i\le M^*$ of these categories, with $M^*<M$,\footnote{We limit the number of preferred categories to $M^*$ to avoid having users who like everything.}
and the set of preferred categories of user $i$ is denoted as $C_i$. As an example, if user $i$ is interested in science and technology, which have category labels 3 and 6, then $C_i=\{3,6\}$ and $m_i=|C_i|=2$.
Each preferred category $c$ of user $i$ is represented by a $D$-dimensional binary taste vector $\mathbf{t}_i^c$, with entries representing the preference for the relative sub-categories.
Specifically, taste vectors have a fixed number ($D_A$) of elements equal one (preferred sub-categories) and all remaining elements equal zero.
The user $i$ in the above example may have for instance $\mathbf{t}_i^{(3)}=(0001100101)$ and $\mathbf{t}_i^{(6)}=(1010000011)$, which correspond to $D=10$ and $D_A=4$.
We make the restriction that any two users cannot have identical taste vectors corresponding to the same category, meaning that there are not identical users in the system.
Summarizing, in the multiple vector model users differ by how many categories they are interested in, by their particular preferred categories and by their specific preferences inside the categories.
We remark that the single vector model used in~\cite{Medo-2009} represents a special case of our multiple vector model---corresponding to $M=1$ and $m_i=1$ $\forall i$.

\begin{figure}[t]
\centering
  \resizebox{0.5\textwidth}{!}{%
  \includegraphics{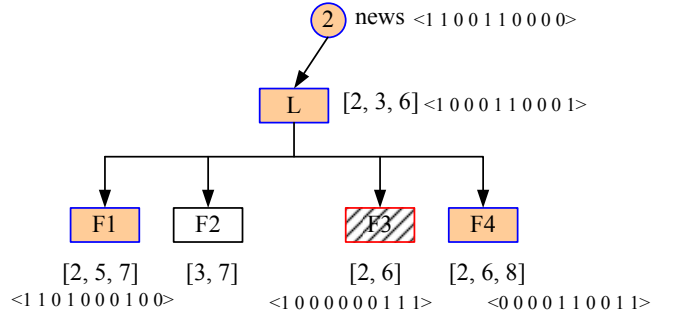}}
\caption{Example of news spreading in the multiple vector model. The numbers inside square brackets are users' favorite categories, whereas the vectors represent the users' tastes and news' attributes in category $c=2$}\label{example}
\end{figure}

Each news $\alpha$ in the system belongs to a single category, hence it is represented by a category label $c$ and a $D$-dimensional attribute vector $\mathbf{a}_\alpha^c$.
Category and attributes of a news are assigned when the news is initially introduced to the system by a user $i$: $c$ is taken at random among $i$'s preferred categories ($c\in C_i$),
and the attribute vector is set identical to $i$'s taste vector corresponding to that category ($\mathbf{a}_\alpha^c\equiv\mathbf{t}_i^c$).
The opinion of a reader $j$ about news $\alpha$ is based on the overlap of the news' attributes with the user's tastes in the category the news belongs to:
\begin{equation}
\label{eq.Ovlp}
\Omega_{j\alpha}=\langle\mathbf{t}_j^c,\mathbf{a}_\alpha^c\rangle
\end{equation}
where $\langle\cdot,\cdot\rangle$ is a scalar product of two vectors and $c$ is the news' category. If $\Omega_{j\alpha}\geq\Delta$ user $j$ likes news $\alpha$ ($e_{j\alpha}=+1$),
otherwise she dislikes it ($e_{j\alpha}=-1$). Here $\Delta$ is the users' approval threshold. However, if the news' category $c$ is not one of $j$'s preferred categories ($c\notin C_j$), then $j$ automatically dislikes the news.
Figure \ref{example} shows an example of the propagation of a news $\alpha$ with $c=2$ and $\mathbf{a}_\alpha^{(2)}=(1100110000)$ in a system with $\Delta=2$.
The news arrives to a certain user $l$ for which $c\in C_l$ and $\Omega_{l\alpha}\geq\Delta$, hence $l$ likes the news and forwards it to her followers.
Among these followers, only $F_2$ does not have $c$ in her preferred categories ($c\notin C_{F_2}$) so she directly dislikes the news.
For $F_3$, $c\in C_{F_3}$ but $\Omega_{F_3\alpha}<\Delta$ so she also dislikes the news. Instead both $F_1$ and $F_4$ at the same time are interested in category $c$
and get an overlap with $\alpha$'s attributes greater or equal than $\Delta$, so they like the news and forward it to their followers.

Simulation runs in discrete time steps. Assuming no a priori information, the starting network configuration is given by randomly assigning $L$ leaders to each user.
Then in each step, an individual user is active with probability $p_A$. When active, the user reads and evaluates the $R$ top-recommended news
she has received and with probability $p_S$ submits a new news. The network of connections is rewired every $u$ time steps.
Parameters values used in all following simulations are given in Table~\ref{parameter}.\footnote{Refer to~\cite{Chen-2011} for a discussion about how the specific choice of parameters influences the simulation results.}

\begin{table}[t]
  \centering
  \caption{List of parameters used in simulations.}\label{parameter}
  \begin{tabular}{l | cccc ccc}
    parameter				& symbol 	& value		\\ \hline	
    Number of users	 		& $U$ 	 	& 3003 		\\
    Number of leaders per user		& $L$ 	 	& 10 		\\
    Total number of categories 		& $M$ 		& 10		\\
    Max. number of preferred categories & $M^*$ 	& 4		\\
    Dimension of taste vectors       	& $D$   	& 14 		\\
    Active elements per vectors       	& $D_A$   	& 6 		\\
    Users' approval threshold  		& $\Delta$ 	& 3		\\
    Probability of being active		& $p_A$ 	& 0.05 		\\
    Probability of submitting a news 	& $p_S$ 	& 0.02 		\\
    Number of news read when active  	&$R$  		& 3		\\
    Damping of recommendation score 	& $\lambda$ 	& 0.9 		\\
    Base similarity for users 		& $s_0$ 	& $10^{-7}$	\\
    Period of the rewiring 		& $u$ 		& 10		
  \end{tabular}
\end{table}

\section{Results}

We now study the described adaptive social recommender system under different definitions of the similarity measure employed.
We use five indices to measure the recommender system's performance and the properties of the leader-follower network:
\begin{itemize}
  \item \emph{Average differences}, the average number of vector elements in which users differ from their leaders: they measure how well the network has adapted to users' tastes and are defined as
$\mbox{a.d.}=\frac{1}{LU}\sum_i\sum_{l\in L_i}\frac{1}{m_l}\\\left(\sum_{c\in C_l\cap C_i}||t_i^c-t_l^c||
+\right.$ $\left.\sum_{c\in C_l\setminus (C_l\cap C_i)}2D_A\right)$.
This definition comes from the following observation. We do not consider the categories that are not preferred by the leader $l$ (also if $i$ has preference for them)
as the news belonging to them cannot be forwarded from $l$ to $i$. Instead for a category preferred by $l$ we distinguish two cases:
if also $i$ has preference for it, we add the relative vector difference; otherwise, we add the maximum difference between two vectors (equal to $2D_A$) as this is the most undesirable scenario---$l$
forwarding news that do not match at all $i$'s preferences.
  \item \emph{Approval fraction}, the ratio of news approvals to all assessments: it tells how often users are satisfied with the news they read and is defined as\\
$\mbox{a.f.}=\sum_{i\alpha}\delta_{e_{i\alpha},1}/\sum_{i\alpha}\delta_{|e_{i\alpha}|,1}$
  \item \emph{Average news' coverage} $\mu=\langle K_\alpha\rangle$, the average number of readers for a news: it measures how broad the news has spread and is defined as
$\mu=\langle\sum_{i}\delta_{|e_{i\alpha}|,1}\rangle_\alpha$
  \item \emph{Coverage heterogeneity} $H=[1+(\sigma_\mu/\mu)^2]^{-1}$
  \item \emph{Fraction of dead ends} (d.e.) or percentage of users with no followers, from which a news cannot propagate further.
\end{itemize}

In addition to the adaptive networks evolving under the different similarity indices already introduced, we also consider a static system
in which the network of connection is artificially constructed to minimize the average differences, i.e. we assume to know users' underlying tastes.

\begin{figure}[t]
\centering
  \resizebox{0.5\textwidth}{!}{%
\includegraphics{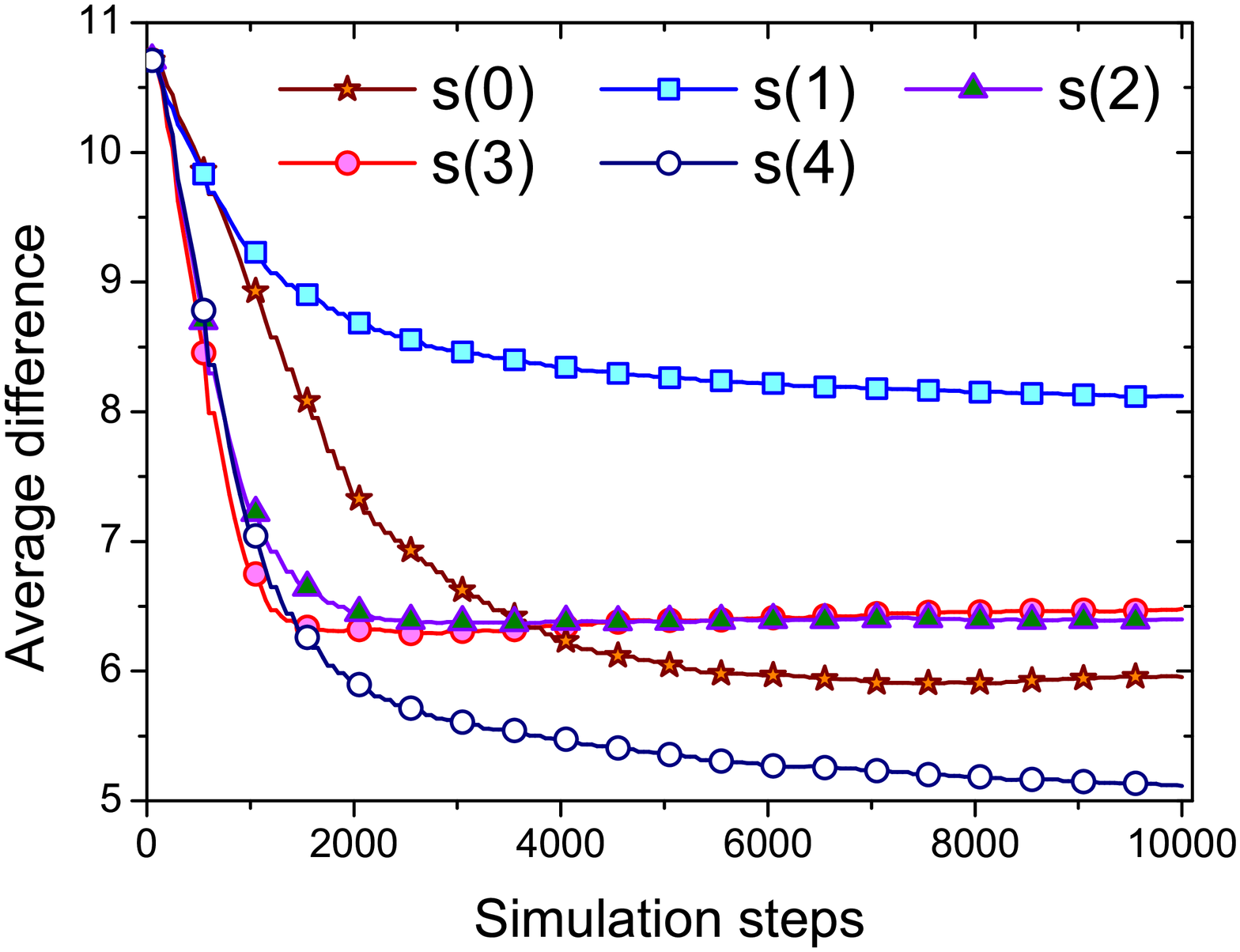}}
\resizebox{0.5\textwidth}{!}{%
\includegraphics{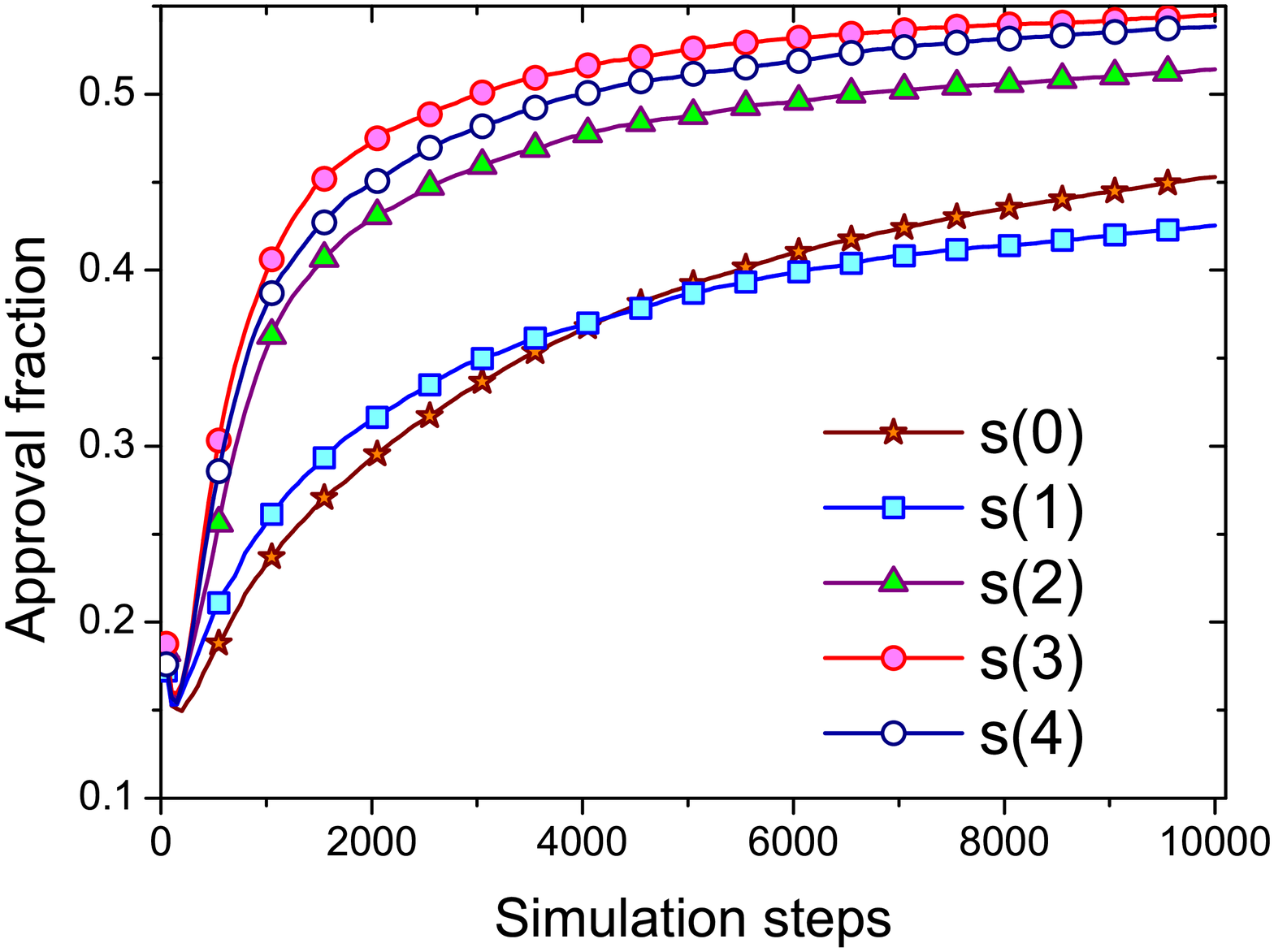}}
\caption{Evolution of average differences (upper panel) and approval fraction (lower panel) in the adaptive system ruled by different definitions of the similarity.
Refer to the next section for the definition of $s^{(4)}$.}\label{multi-adf}
\end{figure}

The evolution of average differences and approval fraction in the system is shown in Figure~\ref{multi-adf}, whereas Table \ref{cons-evol} gives an overview of the results.
We see from Figure~\ref{multi-adf} that after a relatively short transient, the average differences in the network stabilize to a stationary value,
which is notably high for $s^{(0)}$ and $s^{(1)}$, and it is the lowest by construction for the artificial network.
Concerning the approval fraction, we first introduce a reference value of 13.6\% obtained when in our system news are recommended to users at random.
Then, looking at Figure~\ref{multi-adf}, we immediately notice that by using the original similarity measure $s^{(0)}$ the recommender system performs quite poorly---the approval fraction is around $45.3\%$.
This suggests that the similarity as defined in equation (\ref{eqSimO}) may be not appropriate for a system of categories and sub-categories, and for heterogeneous users.
About the other similarity definitions, while $s^{(1)}$ does not perform particularly better, both $s^{(2)}$ and $s^{(3)}$ can significantly improve the approval fraction,
achieving values over $50$ \% which are much closer to the one of the artificial network (reported in Table~\ref{cons-evol}).

\begin{table}[t]
  \centering
  \caption{Summary of the recommender system's performance and of the network's properties:
approval fraction (a.f.), average difference (a.d.), average coverage ($\mu$), coverage heterogeneity ($H$), fraction of dead ends (d.e.) and Shannon information entropy ($I$)
for the artificially constructed network and for the adaptive systems ruled by the various similarity definitions. Values at simulation step \# $10^4$.
Refer to the next section for the definition of $s^{(4)}$ and $I$.}\label{cons-evol}
  \begin{tabular}{l | cc cccc}
			& a.f. & a.d. & $\mu$ & $H$ & d.e. & $I$\\ \hline
    $s^{(0)}$	 	& 45.3 	 & 5.95 & 148.0 & 3.1 & 5.9 &0.577\\
    $s^{(1)}$	    & 42.5   & 8.12 & 92.5 & 3.5 & 5.2 &0.401\\
    $s^{(2)}$   	& 51.4   & 6.40 & 140.5 & 2.2 & 6.4 &0.150\\
    $s^{(3)}$		& 54.5 	 & 6.48 & 145.2 & 2.0 & 0.9 &0.153\\
    $s^{(4)}$		& 53.8 	 & 5.11 & 148.8 & 2.2 & 3.9 &0.472\\
    artificial	 	& 65.5 	 & 3.33 & 149.3 & 4.3 & 58.7 &0.763
  \end{tabular}
\end{table}

\begin{figure}[t]
\centering
\resizebox{0.5\textwidth}{!}{%
\includegraphics{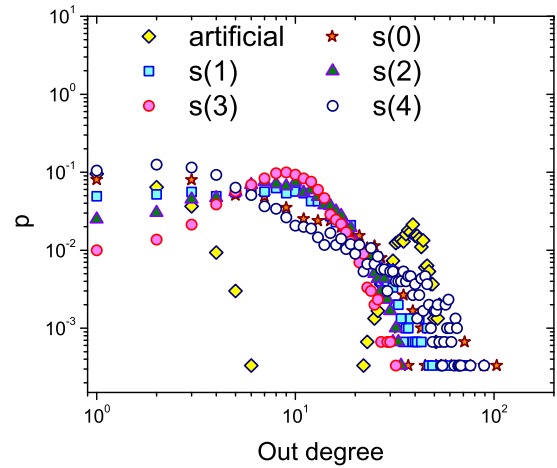}}
\resizebox{0.5\textwidth}{!}{%
\includegraphics{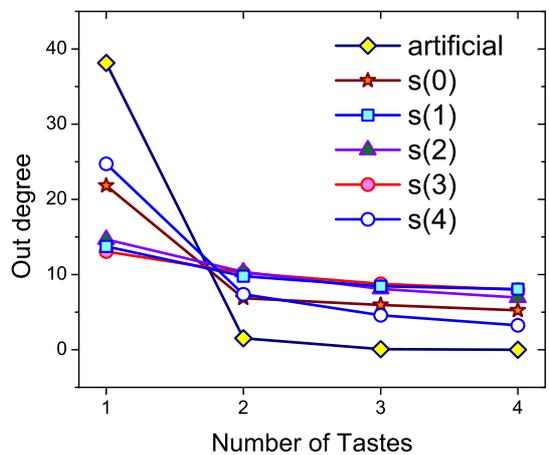}}
\caption{Probability distribution for the number of users' followers (upper panel) and average number of followers as a function of the number of users' preferred categories $m$ (lower panel)
at simulation step \# $10^4$ and for different similarity measures used. Refer to the next section for the definition of $s^{(4)}$.}\label{outdegree}
\end{figure}

We move further by studying additional properties of the leader-follower network. Recalling that the number of leaders per user, $L$, is fixed, but there's no restriction on the number of followers a user can have,
we plot in Figure~\ref{outdegree} the probability distribution of the number of followers and the relation between the number of users' preferred categories and the average number of users' followers.
As shown in the fifth column of Table~\ref{cons-evol}, in the artificial network many users have no followers. This is because in the process of minimizing the average differences users who have many preferred categories
are significantly penalized and are hardly assigned as leaders, becoming in this ways dead ends of the network (see the lower panel of Figure~\ref{outdegree}).
Instead, for the evolving adaptive system under the various definitions of similarity such phenomena is absent: the use of users' assessments does not penalize users with many preferred categories as strongly
as when using taste vectors differences; moreover, the leader selection process is not deterministic, hence also users with wide interests have chances to get some followers.
In these cases the distributions of the number of followers (upper panel of Figure~\ref{outdegree}) are more smooth with respect to the one of the artificial network,
and feature wide tails---users with a few preferred categories are still favored. We remark that the form of these distributions closely resembles the one observed in real systems~\cite{Plosone620648,science286509}.

\section{Recommendation diversity}

Besides providing accurate recommendations, i.e., recommendations for news that are actually liked by users, a good recommender system should also consider the issue of diversity,
by avoiding recommending always the same kind of content. The accuracy-diversity dilemma is a typical feature of recommender system, as often accuracy decreases when diversity improves,
and vice-versa---with few exceptions~\cite{PNAS1074511}.

The result of the previous section is that the highest degree of users' satisfaction is obtained by a network configuration in which highly selective users (with a few preferred categories) are chosen as leaders.
In this situation there is the risk that, for any user, if the few preferred categories of her leaders overlap, then the users will be recommended with news covering only these categories,
while she can still have additional interests---but never receiving news about them, resulting in poor information diversity.
To avoid such undesirable situation, beside the leader-follower similarity, the leader selection process should also account for the similarity among the leaders themselves.
Therefore we introduce another similarity metric:
\begin{equation}
s_{ij}^{(4)}=s_{ij}^{(3)}-\frac{1}{L-1}\sum_{l\in L_i, l\neq j}\frac{|A_j\bigcap A_l|}{|A_j\bigcup A_l|}
\end{equation}
which is based on $s^{(3)}$ (the best performing in accuracy) with an additional term that aims at minimizing the similarity between the candidate leader $j$ and the current leaders of user $i$.
Note that for the second term we use a symmetric Jaccard index (as there is no role difference among leaders), and we do not consider dislikes (see the discussion in the section about the similarity measures).

In order to measure the recommendation diversity in our adaptive system, we use the number of recommendations for each category.
Specifically, if we denote $f_i^c$ as the frequency for which user $i$ reads news belonging to category $c$, we can introduce the standard Shannon information entropy~\cite{BSTJ27379}:
\begin{equation}
I=-\frac{1}{U}\sum_i\sum_{c\in C_i}f_i^c\ln(f_i^c)
\end{equation}
which is maximal when the frequencies are the same (maximum disorder)\footnote{The maximum value of $I$ can be computed as $-\sum_m P(m)\log(m)$, where $P(m)$ is the probability that a users has preference for $m$ categories.}
and zero if each user only reads news of a single category.

\begin{figure}[t]
\centering
\resizebox{0.5\textwidth}{!}{%
\includegraphics{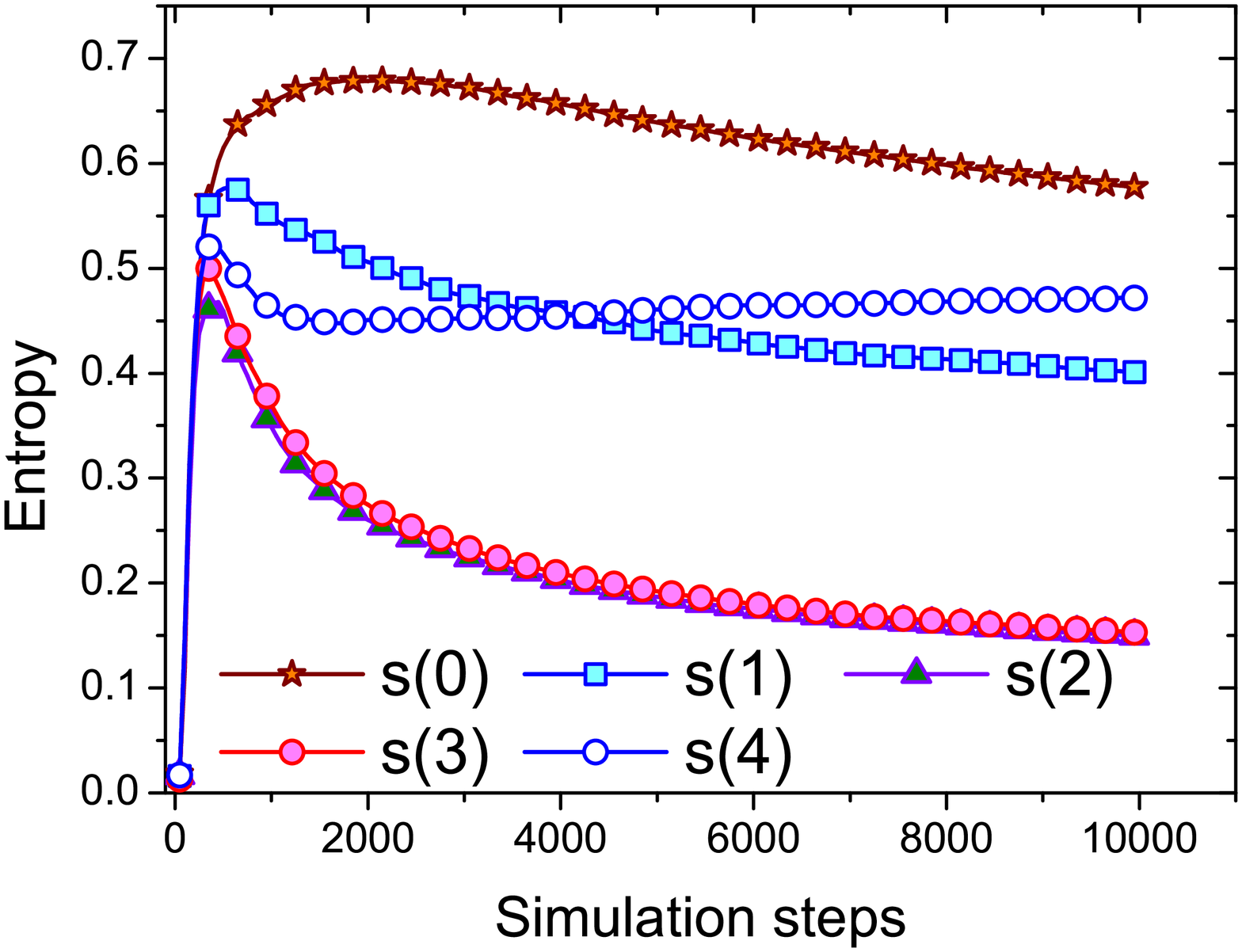}}
\caption{Evolution of the information entropy in the adaptive system ruled by different definitions of the similarity.}\label{entropy}
\resizebox{0.5\textwidth}{!}{%
\includegraphics{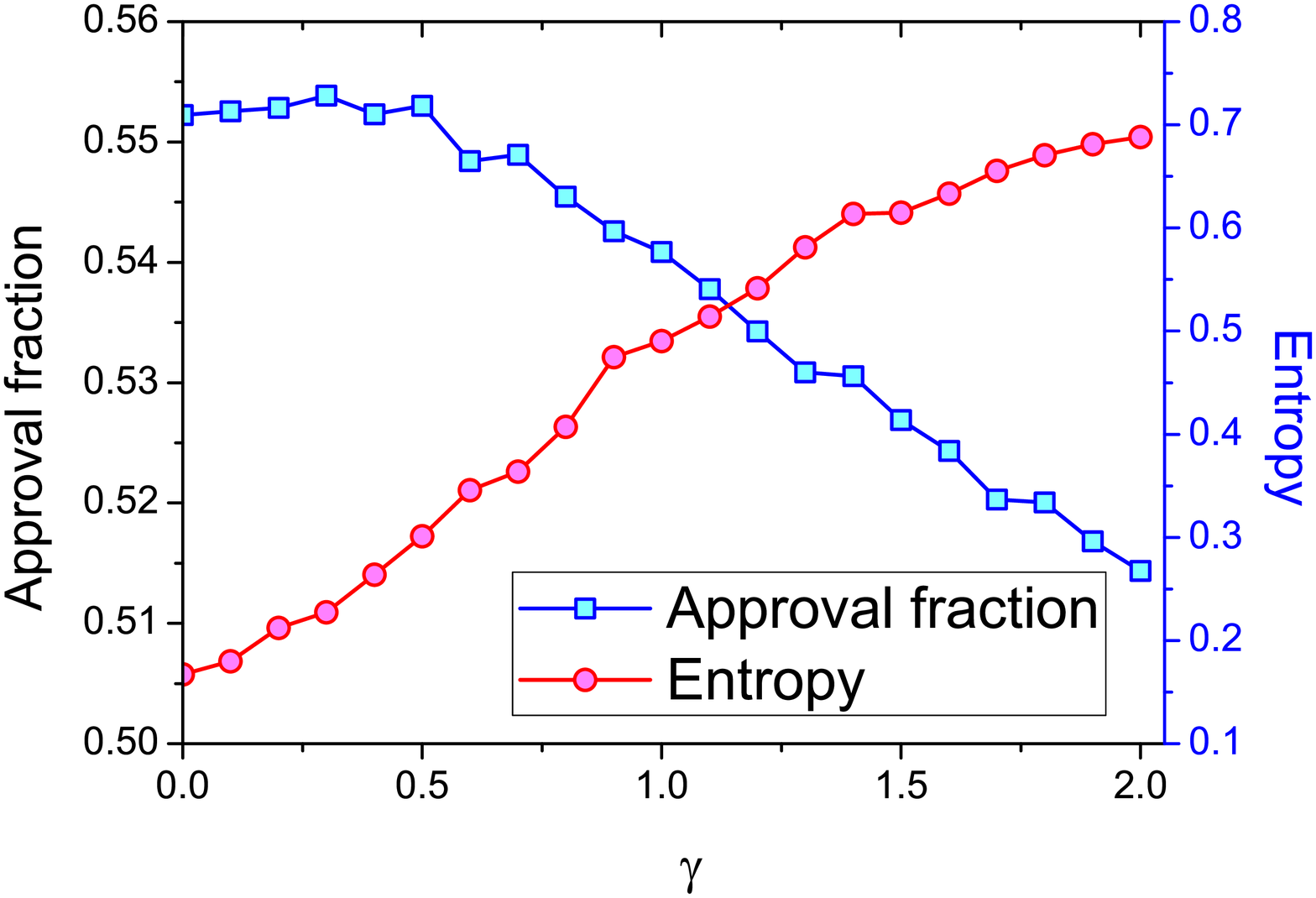}}
\caption{Stationary values (simulation step \# $10^4$) of approval fraction and information entropy in the system when $\tilde{s}_{ij}^{(4)}$ is used, and for different values of the parameter $\gamma$.}\label{gamma}
\end{figure}

The evolution of the information entropy in the adaptive system is shown in Figure~\ref{entropy} and its final values are reported in Table~\ref{cons-evol}.
Comparing Figures~\ref{multi-adf} and~\ref{entropy}, we immediately observe the accuracy-diversity dilemma: $s^{(0)}$, which is the worst performing in approval fraction, achieves the highest diversity,
whereas, $s^{(3)}$, which achieves the highest approval fraction, has the worst performance in diversity. The newly proposed $s^{(4)}$ features a degree of accuracy very similar to the one of $s^{(3)}$,
and at the same time achieves slightly better average differences. However when it comes to diversity, $s^{(4)}$ significantly outperforms $s^{(3)}$.
Summarizing, using $s^{(4)}$ as similarity measure (i.e., minimizing the similarity among leaders), allows to significantly enhance the recommendation diversity while effectively preserving the recommendation accuracy.

The final point we address is to what extent one should consider the similarity among leaders in order to obtain good recommendations. There are two extreme cases here:
considering only the leader-follower similarity as in $s^{(3)}$ results in very low diversity, whereas, if too much weight is given to the second term of $s^{(4)}$ then
the approval fraction may suffer significantly. In order to find the best compromise between accuracy and diversity of our recommendation model,
we introduce a tunable parameter $\gamma$ in the definition of $s^{(4)}$ and obtain:
\begin{equation}
\tilde{s}_{ij}^{(4)}=s_{ij}^{(3)}-\frac{\gamma}{L-1}\sum_{l\in L_i, l\neq j}\frac{|A_j\bigcap A_l|}{|A_j\bigcup A_l|},
\end{equation}
which reduces to $s^{(3)}$ when $\gamma=0$, and to $s_{ij}^{(4)}$ when $\gamma=1$. Clearly, $\gamma$ controls the weight given to the similarity among leaders.

The stationary values of approval fraction and information entropy obtained by using $\tilde{s}_{ij}^{(4)}$ for different values of $\gamma$ are reported in Figure~\ref{gamma}.
We first observe that, in the range of the $\gamma$ values considered, the information entropy increases linearly and significantly with $\gamma$, becoming at the end very close to its maximum value.
The approval fraction shows instead an opposite trend: it decreases with $\gamma$, as expected, although the reduction is only of a few percentage points.
This means that by using $\tilde{s}_{ij}^{(4)}$ it is possible to considerably gain in diversity, at the small cost of slightly reducing accuracy.
More importantly, the approval fraction has an initial plateau---for $\gamma\leq 0.5$, its value remains almost constant. Using a value of $\gamma$ in this region hence allows to obtain higher diversity
of the news users read (up to twice the initial value of $I$), without harming at all the recommendations' accuracy.

\section{Conclusions}

How to deliver the right content to the right user is a fundamental issue in the modern society facing information overload.
Recommender systems represent a possible answer to this problem, and are currently widely-used as information filtering tools.
Recently, the use of social connections to obtain recommendations has emerged, and various adaptive social recommendation models have been proposed by researchers.
Numerical tests of these models often require an agent-based approach, where users and content have to be modeled in a simple yet realistic way.

In this work we studied the social recommendation process within an agent-based framework where users' tastes are modeled by multiple vectors.
Our approach allows to model heterogeneity of users in a rather exhaustive way, while being fairly simple to treat.
We proposed and studied several alternative indices to measure users' taste similarity and build the leader-follower network, and determined the ones for which the system produces more accurate recommendations.
We found that users are more satisfied when their leaders are selective users with a few preferred categories but who are reliable by only forwarding the contents belonging to those categories.
As in such situation there is the risk for users to always get recommended with the same kind of content, we finally discussed the accuracy-diversity dilemma,
and propose additional similarity indices which significantly increase the diversity of the recommendation process without harming its accuracy.


\section*{acknowledgments}
This work was partially supported by the National Natural Science Foundation of China under Grant Nos. 11105025, 61103109 and 60903073,
by the Future and Emerging Technologies programme of the European Commission FP7-COSI-ICT (project QLectives, grant no. 231200) and by the Swiss National Science Foundation (grant no. 200020-121848).

\end{document}